# High-contrast Ultrabroadband Frontend Source for High Intensity Few-Cycle Lasers

A. Jullien[1], X. Chen[1], A. Ricci[1], J. P. Rousseau[1], R. Lopez-Martens[1],
L. P. Ramirez[2], D. Papadopoulos[2,3], A. Pellegrina[2,3], F. Druon[2], P. Georges[2]

[1]*Laboratoire d'Optique Appliquée, ENSTA ParisTech, Ecole Polytechnique, CNRS, 91761 Palaiseau Cedex, France*
[2]*Laboratoire Charles Fabry de l'Institut d'Optique, CNRS, Univ Paris-Sud, 91127 Palaiseau, France*
[3]*Institut de la Lumière Extrême, CNRS, Ecole Polytechnique, ENSTA ParisTech, Institut d'Optique, Univ Paris-Sud, Palaiseau Cedex, France*

**Abstract :** An ultrabroadband seed source for high-power, high-contrast OPCPA systems at 800 nm is presented. The source is based on post compression in a hollow-core fiber followed by crossed polarized waves (XPW) filtering and is capable of delivering 80μJ, 5fs, CEP-stable (0.3rad RMS) pulses with excellent spectral and temporal quality.

High temporal quality few-cycle pulses are needed as seeds for high-power OPCPA type systems to enhance the final contrast [1]. Such ultrashort pulses must feature both high incoherent contrast as well as high spectral quality to minimize the relative intensity of the coherent pedestal and satellite pulses. The preservation of the carrier-envelope phase (CEP) value is also required. XPW generation in $\chi^{(3)}$-anisotropic crystals (BaF2) is a well-known technique leading to significant improvement of the contrast of femtosecond pulses as well as remarkable spectral smoothing [2, 3]. It has been recently demonstrated that XPW generation preserves the CEP [4]. Furthermore, in our previous work, we have shown that this achromatic third-order nonlinear effect is well suited for processes involving ultra-broadband spectra [5]. In this paper, we demonstrate the efficient temporal cleaning of high energy (550 μJ), 5 fs pulses by the XPW filter. On a day-to-day basis 80 μJ pulses with a duration of 5.5 fs and excellent spectral quality (quasi-Gaussian XPW spectrum over 350 nm) are generated. The temporal characterization of these ultrashort XPW pulses shows, for the first time, such an impressive improvement of their coherent and incoherent contrast.

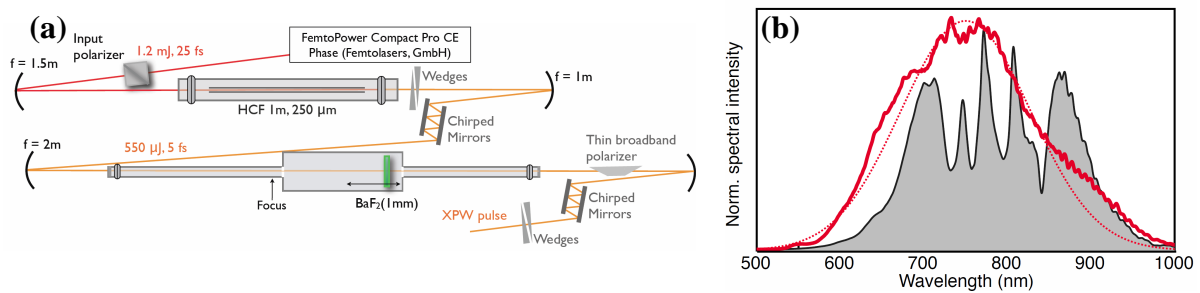

**Fig. 2** : a) Experimental setup, b) Spectra after HCF (grey area), XPW (red line) and its Gaussian fit (red dotted line).

The experimental setup (Fig. 1(a)) includes a commercial 1 kHz CEP-stabilized laser delivering up to 1.5 mJ, 25 fs pulses (Femtolasers, GmbH). To reach the few-cycle regime, about 1.2 mJ, are focused into a hollow-core fiber (HCF) filled with neon and compressed by a set of broadband chirped mirrors and a pair of fused-silica wedges. The obtained 550 μJ, 5 fs compressed pulses are then seeded into the XPW filter. The beam is focused by a f=2 m mirror into a vacuum chamber to avoid nonlinear pulse distortion in air. The 1 mm thick nonlinear crystal (BaF2, [011] crystallographic orientation) is placed out of focus to reach the adequate peak intensity for XPW generation. The XPW signal is then discriminated by a broadband thin film polarizer with low dispersion and good transmission over a broad spectral bandwidth, allowing the compression of the XPW pulse. However, its polarizing efficiency is limited so that, with an input polarizer before the HCF, the setup extinction ratio is $10^{-3}$-$10^{-2}$:1. The XPW pulse is sent to a second chirped mirror compressor for temporal characterization.

According to our previous investigations [5], accurate tuning of the input pulse dispersion ($\pm$ 4 fs$^2$) allows fine optimization of the XPW process. The filter transmission efficiency is then 15% and 80 µJ pulses are generated. Under the specific compression conditions, the fast electronic response of the XPW process enables efficient spectral cleaning: the XPW output spectrum exhibits a smooth, near-Gaussian shape over 350 nm (Fig. 1(b)). This excellent spectral property is a consequence of the temporal quality enhancement occurring during the process. Fig. 2 (a) shows a comparison of the FROG traces measured before and after XPW. For the XPW pulses, temporal side lobes and sharp features typical of few-cycle pulses are decreased and the spectro-temporal energy distribution is balanced and homogeneous. The measured duration of the XPW pulse is 5.5 fs and the coherent background is clearly reduced on the femtosecond time-scale. After calculated additional propagation in glass (~100 µm) to remove the low residual quadratic phase, we found that the main peak contains 91% of the energy after XPW, for only 79% after the HCF.

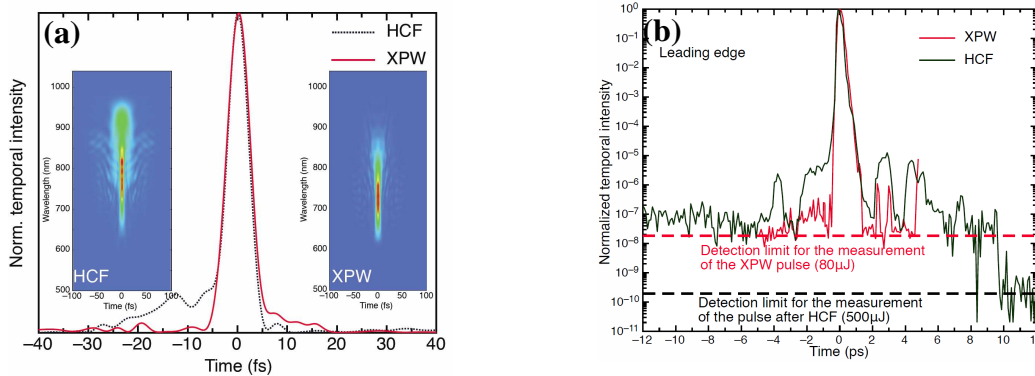

**Fig. 2** : a) FROG measurements of the compressed few-cycle pulses after HCF and after XPW, b) Temporal contrast of the few-cycle pulses measured by a third-order high-dynamical cross-correlator before and after the XPW filter.

A rough estimation of the improvement of the temporal contrast on a picosecond time-scale is provided by a third-order high-dynamical cross-correlator measurement (Fig. 2(b)). The device is not designed for ultra-broadband spectra and consequently over-estimates the background intensity relative to the main peak. Furthermore, due to the low energy of the XPW signal, the dynamic range of the device is limited to 8 orders of magnitude in the XPW case. However, the measurement indicates a contrast improvement of at least 2 orders of magnitude (consistent with the extinction ratio of the setup) and for the first time, a 5 fs pulse is shown to present a steep rising edge ($<10^{-7}$ at t = -2 ps) and a temporal contrast better than $10^{-8}$ (detection limited). Finally, using a home-made f-to-2f interferometer, we verified the CEP stability after these two consecutive nonlinear stages (HCF+XPW) measured to about 0.3 rad RMS. The CEP stability could be further improved by simple protection of the whole experimental setup to avoid air fluctuations.

**Ackowledgements:** The authors gratefully acknowledge P. Monot for the high-dynamical contrast measurements.